\documentstyle[proceedings,psfig]{heliosymp} 

\makeatletter
\def\astropbibitem{\@lbibitem}
\def\@lbibitem#1#2#3{\item[\hfill]\if@filesw 
      { \def\protect##1{\string ##1\space}\immediate
        \write\@auxout{\string\astropbibcite{#3}{#1}{#2}}\fi\ignorespaces}}
\def\astropbibcite#1#2#3{\global\@namedef{b@#1}{#2\ #3} %without comma 
			\global\@namedef{newb@#1}{#2\ (#3}
			\global\@namedef{nameb@#1}{#2}
			\global\@namedef{yearb@#1}{#3}}
\let\citation\@gobble

\def\cite{\@ifnextchar [{\@tempswatrue\@citex}{\@tempswafalse\@citex[]}}
\def\citeone{\@ifnextchar [{\@tempswatrue\@newcitex}
			   {\@tempswafalse\@newcitex[]}}
\def\citename{\@ifnextchar [{\@tempswatrue\@namecitex}
			   {\@tempswafalse\@namecitex[]}}
\def\citeyear{\@ifnextchar [{\@tempswatrue\@yearcitex}
			   {\@tempswafalse\@yearcitex[]}}
\def\@citex[#1]#2{\if@filesw\immediate\write\@auxout{\string\citation{#2}}\fi
  \def\@citea{}\@cite{\@for\@citeb:=#2\do
     {\@citea\def\@citea{; }\@ifundefined
       {b@\@citeb}{{\bf ?}\@warning
       {Citation `\@citeb' on page \thepage \space undefined}}%
{\csname b@\@citeb\endcsname}}}{#1}}
\def\@newcitex[#1]#2{\if@filesw\immediate\write\@auxout{\string\citation{#2}}\fi
  \def\@newcitea{}\@newcite{\@for\@newciteb:=#2\do
     {\@newcitea\def\@newcitea{; }\@ifundefined
       {newb@\@newciteb}{{\bf ? (?}\@warning
       {Citation `\@newciteb' on page \thepage \space undefined}}%
{\csname newb@\@newciteb\endcsname}}}{#1}}

\def\@namecitex[#1]#2{\if@filesw\immediate\write\@auxout{\string\citation{#2}}\fi
  \def\@namecitea{}\@namecite{\@for\@nameciteb:=#2\do
     {\@namecitea\def\@namecitea{; }\@ifundefined
       {nameb@\@nameciteb}{{\bf ? (?}\@warning
       {Citation `\@nameciteb' on page \thepage \space undefined}}%
{\csname nameb@\@nameciteb\endcsname}}}{#1}}

\def\@yearcitex[#1]#2{\if@filesw\immediate\write\@auxout{\string\citation{#2}}\fi
  \def\@yearcitea{}\@yearcite{\@for\@yearciteb:=#2\do
     {\@yearcitea\def\@yearcitea{; }\@ifundefined
       {yearb@\@yearciteb}{{\bf ? (?}\@warning
       {Citation `\@yearciteb' on page \thepage \space undefined}}%
{\csname yearb@\@yearciteb\endcsname}}}{#1}}

\let\bibdata=\@gobble
\let\bibstyle=\@gobble

\def\bibliography#1{\if@filesw\immediate\write\@auxout{\string\bibdata{#1}}\fi
  \@input{\jobname.bbl}}

\def\bibliographystyle#1{\if@filesw\immediate\write\@auxout
    {\string\bibstyle{#1}}\fi}
\def\nocite#1{\@bsphack
  \if@filesw\immediate\write\@auxout{\string\citation{#1}}\fi
  \@esphack}
\def\@cite#1#2{({#1\if@tempswa ; #2\fi})}
\def\@newcite#1#2{{#1\if@tempswa ; #2\fi})}
\def\@namecite#1#2{#1}
\def\@yearcite#1#2{#1}
\newcommand{\citebare}[1]{{\citename{#1}\ \citeyear{#1}}} %without comma 
\newcommand{\citeeg}[1]{{(e.g., {}\citebare{#1})}}

\newcommand{\citeabbare}[2]{{\citename{#1}\ \citeyear{#1},b}\nocite{#2}}

\def\thebibliography#1{\section*{References}\list
 {}{\setlength\labelwidth{1.4em}\leftmargin\labelwidth
 \setlength\parsep{0pt}\setlength\itemsep{0pt}
 \setlength{\itemindent}{-\leftmargin}
 \usecounter{enumi}}
 \def\newblock{\hskip .11em plus .33em minus -.07em}
 \sloppy
 \sfcode`\.=1000\relax}

\makeatother

\newcommand{\eBoo}{\mbox{$\eta$~Boo}}
\newcommand{\aCenA}{\mbox{$\alpha$~Cen~A}}

\begin{opening}
\title{A search for solar-like oscillations in \aCenA}

\author{T.R. Bedding}
\institute{School of Physics, University of Sydney 2006, Australia}

\author{H. Kjeldsen} 

\institute{Teoretisk Astrofysik Center, Danmarks Grundforskningsfond,
\protect\\ Aarhus University, DK-8000 Aarhus~C, Denmark}

\author{S. Frandsen} 

\author{T.H. Dall} 

\institute{Institut for Fysik og Astronomi, Aarhus Universitet, DK-8000
Aarhus~C, Denmark}

\end{opening}

\runningtitle{Solar-like oscillations}

\begin{document}

{\fontsize{10pt}{11pt}\selectfont % DON't REMOVE THIS LINE !
% Note that this command needs the bracket before the \end{document}

\begin{abstract}
We have been using a new method to search for solar-like oscillations that
involves measuring temperature changes via their effect on the equivalent
widths of the Balmer hydrogen lines.  We observed \aCenA\ over six nights
in 1995 with the 3.9-metre Anglo-Australian Telescope and the European
Southern Observatory's 3.6-metre telescope in Chile.  Oscillations were not
detected, with an upper limit only slightly higher than the expected
signal.
\end{abstract}

\section{Introduction}

Many attempts have been made to detect stellar analogues of the solar
five-minute oscillations.  As with helioseismology, it is hoped that the
measurement of oscillation frequencies in other stars will place important
constraints on stellar model parameters and provide a strong test of
evolutionary theory.  However, despite several claims in the literature, it
is fair to say that there has been no unambiguous detection of solar-like
oscillations in any star except the Sun (see reviews by \citebare{B+G94};
\citebare{K+B95}; \citebare{B+K98}).

We have been using a new method to search for solar-like oscillations that
involves measuring temperature changes via their effect on the equivalent
widths of the Balmer hydrogen lines.  We found strong evidence for
solar-like oscillations in the G~subgiant \eBoo\ \cite{KBV95,B+K95}, with
frequency splittings that were later found to agree with theoretical models
(\citeabbare{ChDBH95}{ChDBK95}; \citebare{Gu+D96}).  Since then, the
improved luminosity estimate for \eBoo{} from Hipparcos measurements has
given even better agreement \cite{BKChD98}.  However, a search for velocity
oscillations in \eBoo{} by \citeone{BKK97} failed to detect a signal,
setting limits at a level below the value expected on the basis of the
\citename{KBV95} result.  More recently, Brown et al.\ (private
communication) have obtained a larger set of observations which they are
currently processing.

\section{Observations of \aCenA}

We chose \eBoo\ as the first target for the equivalent-width method because
this star was expected to have an oscillation amplitude about five times
greater than the Sun.  This turned out to be the case (assuming the
detection is real).  We then turned to \aCenA, a more challenging target
because of its smaller expected oscillation amplitude (comparable to solar;
\citebare{BKR96}).  Being a near twin of the Sun and extremely nearby, this
star is an obvious target for detecting oscillations \citeeg{BCW94}.
Previous attempts to detect oscillations using Doppler methods were
reviewed by \citeone{K+B95} and include two claimed detections at
amplitudes 4--6 times greater than solar \cite{GGF86,PBvH92} and two
negative results at amplitudes about 2--3 times solar \cite{B+G90,E+C95}.

We observed \aCenA\ over six nights in April 1995 from two sites: 
\begin{itemize}

\item at Siding Spring Observatory in Australia, HK and TRB used the
3.9-metre Anglo-Australian Telescope with a coud\'e echelle spectrograph
(UCLES).  We recorded three orders centred at H$\alpha$ and three orders at
H$\beta$.  The weather was about 85\% clear.

\item at La Silla in Chile, SF and THD used the European Southern
Observatory's 3.6-metre telescope with a Cassegrain echelle spectrograph
(CASPEC).  We recorded three orders centred at H$\alpha$.  The weather was
100\% clear.

\end{itemize}

\section{Results and simulations}

Data processing of the 20,000 spectra was carried out by HK using the
method outlined in \citeone{B+K98}.  The power spectrum of the resulting
time series of equivalent-width measurements is shown in
Fig.~\ref{fig.acen}.  No obvious excess of power is seen -- note that
earlier reports of a positive detection were premature \cite{KFB96,Fra97}.
The average noise level in the amplitude spectrum (square root of power) is
4.7\,ppm, which is somewhat higher than expected purely from photon noise.
One extra noise source arises from wavelength-dependent fluctuations in the
continuum, which appear to arise from a colour term in the scintillation
\cite{JPP76,DLM97b}.

\begin{figure}[p]
\centerline{\psfig{figure=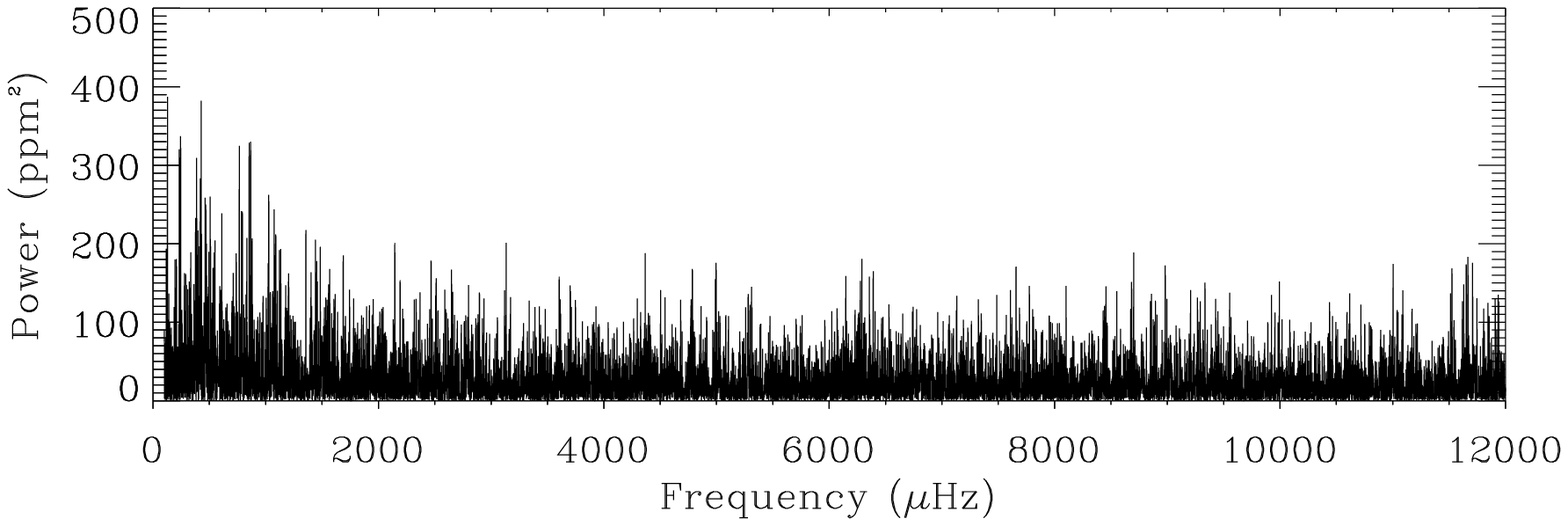,width=\the\hsize}}
\centerline{\psfig{figure=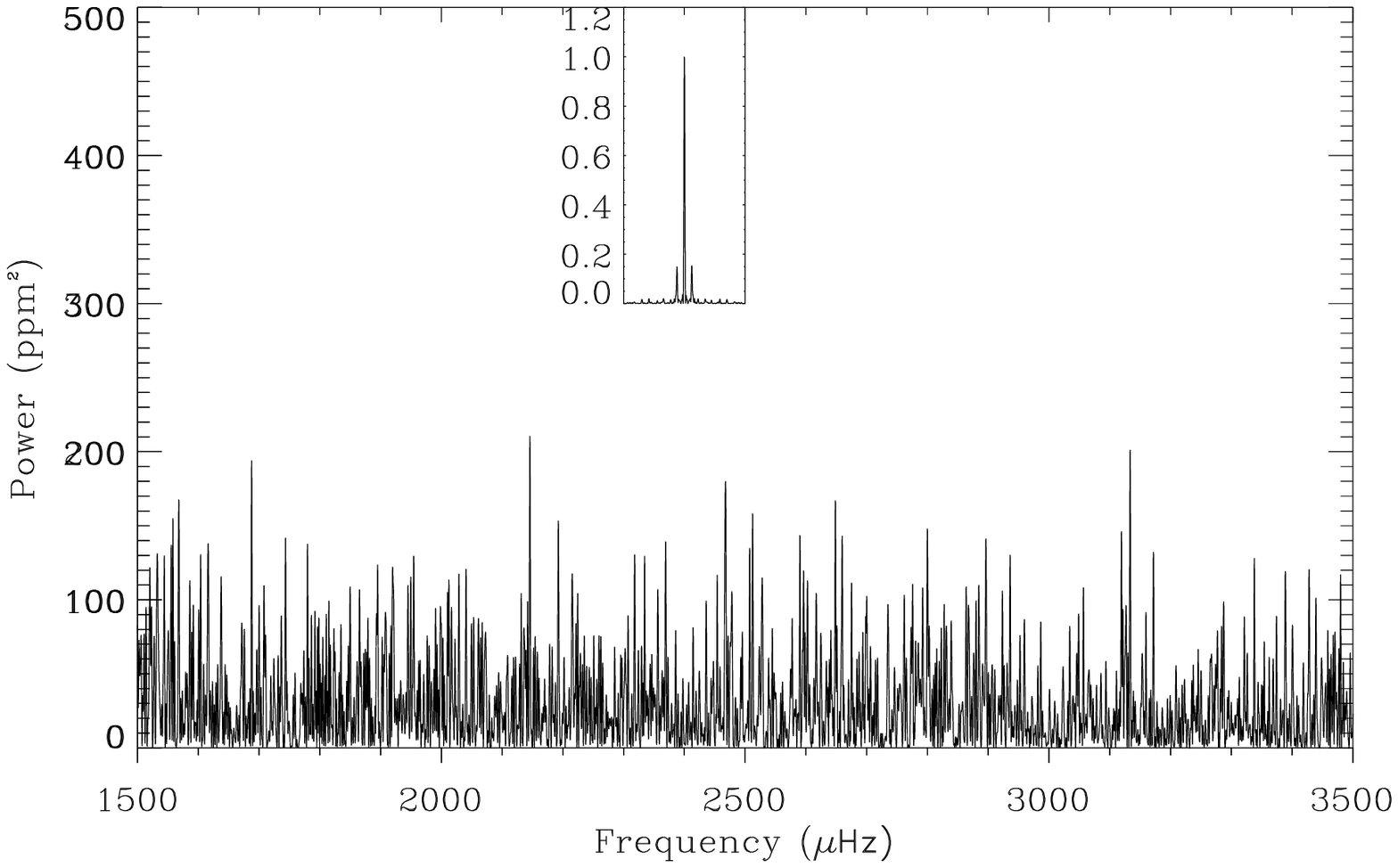,width=\the\hsize}}
\caption[]{\label{fig.acen}Power spectrum of equivalent-width observations
of \aCenA.  The lower figure is a close-up of the region where signal would
be expected, and the inset shows the power spectrum of the window
function.}
\end{figure}

The strongest oscillation modes in \aCenA, as measured in H$\alpha$
equivalent width, are expected to have amplitudes of about 8\,ppm, while
the solar peak amplitude is about 6\,ppm \cite{BKR96}.  

To set an upper limit on oscillation amplitudes from our observations, we
have generated simulated time series consisting of artificial signal plus
noise.  Each simulated series had exactly the same sampling function and
allocation of statistical weights as the real data.  The injected signal
contained sinusoids at the frequencies calculated by \citeone{ECD92},
modulated by a broad solar-like envelope centred at 2.3\,mHz (which is the
expected frequency of maximum mode power -- \citebare{K+B95}).  In each
simulation, the phases of the oscillation modes were chosen at random and
the amplitudes were randomized about their average values.  All these
characteristics were chosen to imitate as closely as possible the
stochastic nature of oscillations in the Sun.  Before calculating the power
spectrum, we added normally-distributed noise to the time series, so as to
produce a noise level in the amplitude spectrum of 4.7\,ppm (consistent
with the actual data).

Some results are shown in Fig.~\ref{fig.sim}.  The top panel shows that a
signal with an amplitude of 12\,ppm would be easily detectable in our data.
It is interesting to note that some of the signal peaks in this simulated
power spectrum have been strengthened significantly by constructive
interference with noise peaks.  For example, a signal peak of 12\,ppm which
happens to be in phase with a 2-$\sigma$ noise peak ($2\times4.7$\,ppm)
will produce a peak in power of 450\,ppm$^2$.  This illustrates the point
made by \citeone[Appendix A.2]{K+B95}: the effects of noise must be taken
into account when estimating the amplitude of a signal.

The next four panels in Fig.~\ref{fig.sim} show simulations in which the
strongest modes had amplitudes of 8\,ppm, the value expected for \aCenA.
In some cases, excess power is seen and there is perhaps some hint of
regularly spaced peaks.  However, the actual data (bottom panel) are
clearly consistent with an 8\,ppm signal and we conclude that the
observations did not have sufficient sensitivity to detect a signal of this
strength.  We can probably set an upper limit of about 10\,ppm.

\begin{figure}[p]
\centerline{\psfig{figure=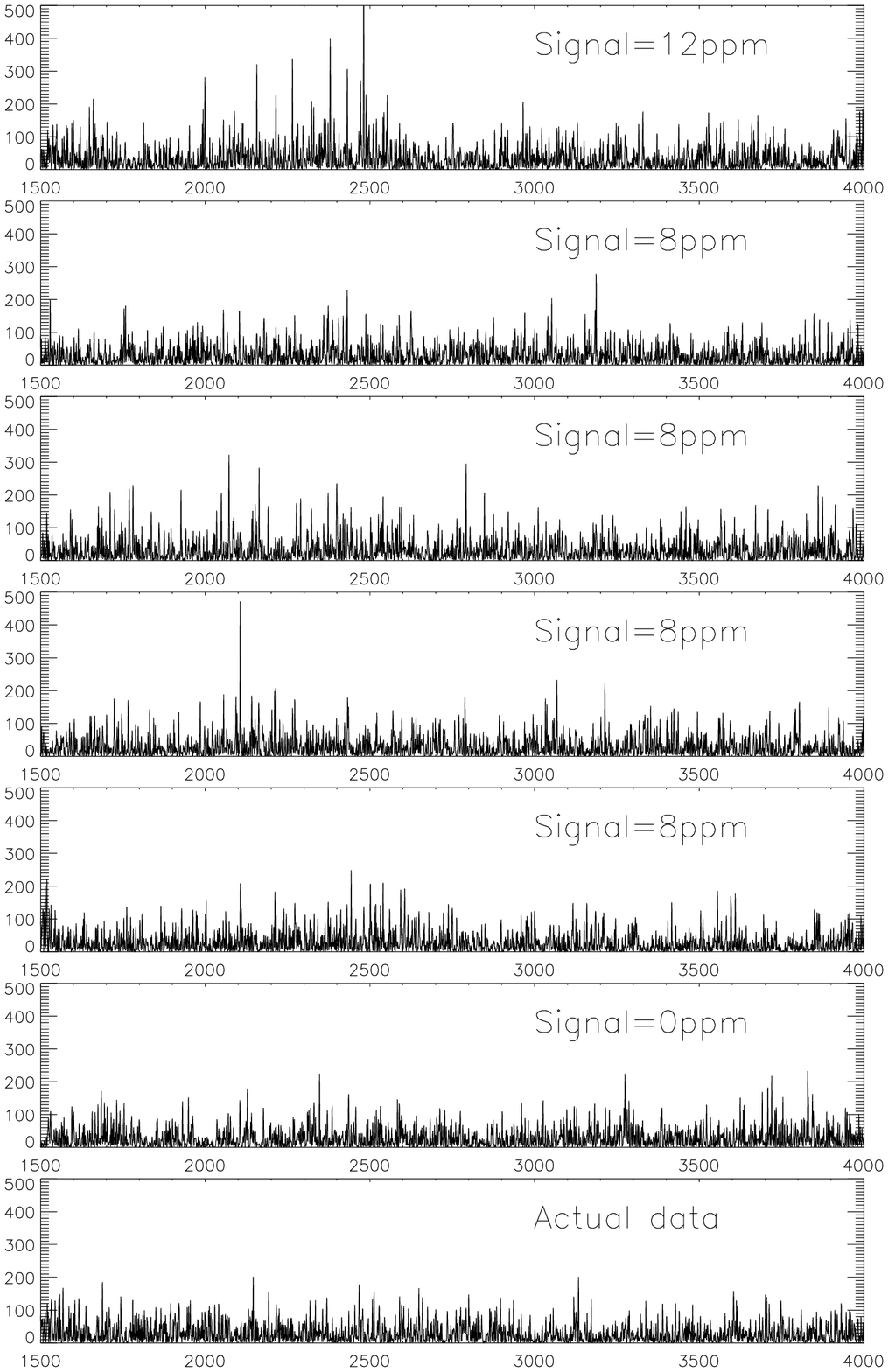,width=12cm}}
\caption[]{\label{fig.sim}Simulated power spectra, using the same sampling
times and data weights as the actual observations.}
\end{figure}

In summary, our results rule out oscillations at a level slightly less than
twice solar, making \aCenA\ the most stable known extra-solar star.  The
corresponding upper limit in velocity is about 50\,cm/s.

\subsection*{Acknowledgements}

The observations would have been impossible without the excellent support
we received from staff at both observatories.  We are especially grateful
to Roy Antaw, Bob Dean, Sean Ryan, John Stevenson and Gordon Shafer at the
AAO and to Luca Pasquini, Peter Sinclaire and Nicolas Haddad at ESO\@.  We
also thank both committees (ATAC and OPC) for allocating telescope time and
the AAO Director for granting the sixth AAT night.  This work was supported
financially by the Australian Research Council and by the Danish National
Research Foundation through its establishment of the Theoretical
Astrophysics Center.

\end{document}